\documentclass[twocolumn,showpacs,prb,superscriptaddress]{revtex4}

\usepackage{graphicx}
\usepackage{tabularx}
\usepackage{epsfig}
\usepackage{amsmath}
\usepackage{amssymb}

\begin{document}

\title{Effective critical  behavior of the two-dimensional Ising spin 
glass with bimodal interactions}

\author{Helmut G.~Katzgraber}
\affiliation{Theoretische Physik, ETH Z\"urich, CH-8093 Z\"urich, 
Switzerland}

\author{L.~W.~Lee}
\affiliation{Department of Physics, University of California,
Santa Cruz, California 95064, USA}

\author{I.~A.~Campbell}
\affiliation{Laboratoire des Collo\"ides, Verres et Nanomateriaux, 
Universit\'e Montpellier II, 34095 Montpellier, France}

\date{\today}
\begin{abstract}

Monte Carlo data of the two-dimensional Ising spin glass with
bimodal interactions are presented with the aim of understanding the
low-temperature physics of the model. An analysis of the specific
heat, spin-glass susceptibility, finite-size correlation length,
and the Binder ratio is performed to try to verify a recent proposal
in which for large system sizes and finite but low temperatures the
effective critical exponents are identical to the critical exponents of
the two-dimensional Ising spin glass with Gaussian interactions. Our
results show that with present system sizes the recently proposed
scenario in which the two-dimensional Ising spin glass with bimodally
distributed interactions is in the same universality class as the
model with Gaussian-distributed disorder at low but finite temperatures
cannot be reliably proven.

\end{abstract}

\maketitle

\section{Introduction}
\label{sec:introduction}

The two-dimensional Ising spin glass\cite{binder:86}
has been the subject of numerous numerical
studies\cite{wang:88,saul:93,houdayer:01,lukic:04,katzgraber:05b}
ever since its introduction by Edwards and Anderson in
1975.\cite{edwards:75} Despite the fact that this canonical model only
orders at zero temperature, its popularity can be ascribed mainly
to its ease of implementation and simplicity. Yet it has proven to
be extremely difficult to establish if the model orders at zero or
finite temperature and what the exact values of the critical exponents
are. Today it is well recognized that the model only orders at zero
temperature.\cite{hartmann:01a} In addition, the critical exponents
at $T = 0$ are known to good precision.\cite{hartmann:01a}

When the interactions between the spins are drawn from a Gaussian
distribution with zero mean, the system has a unique ground
state (up to a global symmetry). This in turn implies that the
critical exponent of the correlation function has to be zero,
i.e., $\eta = 2 - d = 0$, where $d = 2$ is the space dimension.
In addition, extensive zero-temperature domain-wall computations
have established that the domain-wall stiffness exponent is $\theta =
-0.282(2)$.\cite{rieger:96,hartmann:01a} Because at a zero-temperature
transition the critical exponent of the correlation length $\nu$
is related to $\theta$ via $\nu = -1/\theta$, it follows that $\nu
= 3.54(2)$.

The ground state of the model with bimodally distributed random bonds,
on the other hand, is strongly degenerate\cite{blackman:98,lukic:04}
with a finite residual entropy at $T = 0$. Domain-wall stiffness
measurements at zero temperature show an exponent $\theta =
0$\cite{hartmann:01a} (exponential scaling) although with significant
corrections to scaling up to linear system sizes $L \approx 100$
under periodic/free boundary conditions.  Direct measurements
of the spin-spin correlation function $G(r)$ at $T=0$ indicate
a critical exponent $\eta \approx 0.14$.\cite{poulter:05} An
extrapolation from finite temperatures\cite{katzgraber:05b}
yielded to high precision $\eta = 0.138(5)$.  Other
estimates of $\eta$ have given positive values in the range
$0.14$--$0.40$.\cite{mcmillan:83,bhatt:88,wang:88,saul:93,blackman:98,amoruso:06a}

Judging from their zero-temperature (critical) properties the two
versions of the model thus are in two different universality classes
(see Table \ref{tab:exp}). However, for low but finite temperatures
the situation is less clear-cut. For the model with bimodal disorder,
when a sample is in a ground state, turning over one individual spin
either leaves the total energy of the sample unchanged or increases it
by $4J$ or $8J$, where $J$ is the characteristic energy of the model
(see below). This suggests immediately that there is an energy gap
of $AJ$ between the ground state and the first excited state, with
$A=4$. It has been argued however that the true elementary excitations
are not single spins but more complex objects, leading to an effective
gap of $\sim 2J$.\cite{wang:88,houdayer:01,lukic:04,wang:05} Because
of the gapped excitation spectrum we expect a crossover temperature
$T^{*}(L)$, which separates a critical behavior, in accordance with
the aforementioned zero-temperature critical point and a critical
behavior which would resemble a model with continuous interactions. If
the ground-state degeneracy is given by $N_0(L)$ and the excited
states have a degeneracy $N_1(L)$, $N_2(L)$, \ldots, for $N_0(L)
\gg N_1(L)\exp(-4J/T)$ the system will spend almost all of its time
in the ground state and equilibrium properties will be essentially
those of the ground-state manifold, whereas for $N_1(L)\exp(-4J/T)
\gg N_0(L)$ the system will stay in the quantized series of excited
states and the properties of the system can be expected to resemble
those of a system with no gap in the energy spectrum. Note that this
argument is slightly oversimplified as the effects of higher excited
states on $T^{*}(L)$ are not taken into account, yet we expect their
contributions to be small.\cite{fisch:06} One obtains to lowest order
$T^{*}(L) \sim 4J/\ln[N_1(L)/N_0(L)]$. The ratio $N_1(L)/N_0(L)$
with a gap $4J$ has been estimated in Refs.~\onlinecite{lukic:04}
and \onlinecite{fisch:06} and shows that $T^{*}(L)$ drops gradually
as $L$ increases. Our data presented below confirm this behavior.

It has been strongly argued\cite{joerg:06a} that in the limit $T >
T^{*}(L)$ and with $L \rightarrow \infty$ [meaning $T^{*}(L)$ tending
to zero but never reaching zero] the model with bimodal disorder
has effective critical exponents {\em identical} to the critical
exponents of the model with Gaussian-distributed disorder, so that
the two models can be considered as being in the same universality
class except for the singular behavior of the model with bimodally
distributed disorder at $T = 0$.

We present the results of Monte Carlo simulations of the two-dimensional
Ising spin glass with bimodally distributed disorder on
system sizes larger than those used in Refs.~\onlinecite{lukic:04} and
\onlinecite{joerg:06a}. Our results show that with current algorithms
and computer power the data do not provide definitive limiting values
for the critical exponents of the model, although power-law scaling
seems plausible for finite but nonzero temperatures.  Therefore the
claim that the aforementioned model is in the same universality class
at finite but nonzero temperatures as the model with Gaussian, gap
$1/4$, or diluted interactions\cite{joerg:06a} remains to be proven.

The paper is structured as follows: In Sec.~\ref{sec:model} we
introduce the model, numerical method, and observables, and discuss
different finite-size scaling relations. In Sec.~\ref{sec:previous}
we summarize previous results on the two-dimensional Ising spin glass
with bimodally distributed disorder. Results on the different critical
exponents are presented in Sec.~\ref{sec:crit} and a finite-size
scaling analysis of the data is presented in Sec.~\ref{sec:fss}.

\section{Model, Observables, and finite-size scaling relations}
\label{sec:model}

The Hamiltonian of the two-dimensional Ising spin glass is given by
\begin{equation}
{\cal H} = - \sum_{\langle i,j\rangle} J_{ij} S_i S_j .
\label{eq:ham}
\end{equation}
$S_i= \pm 1$ represent Ising spins and the sum is over nearest neighbors 
on a square lattice with periodic boundary conditions. The interactions 
$J_{ij} \in \{\pm J\}$ (here $J = 1$)
are bimodally distributed. For the Monte Carlo 
simulations we use a combination of single-spin flips, exchange Monte Carlo 
updates,\cite{hukushima:96,marinari:98b} and rejection-free cluster 
moves\cite{houdayer:01} to speed up equilibration. Equilibration of the 
method is tested by performing a logarithmic data binning of all 
observables, and we require that the last three bins agree within error 
bars and are independent of the number of Monte Carlo sweeps 
$N_{\mathrm{sweep}}$. The parameters of the simulation are listed in Table 
\ref{simparams}.
\begin{table}
\caption{
Parameters of the simulations. $N_{\mathrm{samp}}$ represents the number 
of disorder realizations computed; $N_{\mathrm{sweep}}$ is the total 
number of Monte Carlo sweeps of the $2 N_T$ replicas for a single sample. 
$N_T$ is the number of temperatures in the exchange Monte Carlo method and 
$T_{\rm min}$ represents the lowest temperature simulated. (For $L = 128$ 
no data for the specific heat has been generated.)
\label{simparams}
}
\begin{tabular*}{\columnwidth}{@{\extracolsep{\fill}} c r r r l }
\hline
\hline
$L$  &  $N_{\mathrm{samp}}$  & $N_{\mathrm{sweep}}$ & $T_{\rm min}$ & $N_{\rm T}$  \\
\hline
 32 & $ 5000 $ & $2.0 \times 10^6$ & 0.050 & 20 \\
 48 & $ 1000 $ & $2.0 \times 10^6$ & 0.050 & 20 \\
 64 & $  500 $ & $4.2 \times 10^6$ & 0.200 & 39 \\
 96 & $  609 $ & $6.5 \times 10^6$ & 0.200 & 63 \\
128 & $  420 $ & $2.0 \times 10^6$ & 0.396 & 50 \\
\hline
\hline
\end{tabular*}
\end{table}

The second-moment finite-size correlation 
length\cite{cooper:82,kim:94,palassini:99b,ballesteros:00,lee:03,katzgraber:04} 
$\xi_L$ is given by
\begin{equation}
\xi_L = \frac{1}{2 \sin (|{\bf k}_\mathrm{min}|/2)}
\left[\frac{\chi_{\mathrm{SG}}(0)}
{\chi_{\mathrm{SG}}({\bf k}_\mathrm{min})} - 1 \right]^{1/2} \; ,
\label{eq:xiL}
\end{equation}
where ${\bf k}_\mathrm{min} = (2\pi/L, 0)$ is the smallest nonzero
wave vector, and $\chi_{\mathrm{SG}}({\bf k})$ is the  wave-vector-dependent
spin-glass susceptibility,
\begin{equation}
\chi_{\mathrm{SG}}({\bf k}) = \frac{1}{N} \sum_{i,j} [\langle S_i S_j
\rangle^2 ]_{\rm av} e^{i {\bf k}\cdot({\bf R}_i - {\bf R}_j) } \; .
\label{eq:chik}
\end{equation}
In the previous equation $[\cdots]_{\rm av}$ represents a disorder average and
$\langle \cdots \rangle$ a thermal average. The finite-size correlation length
is expected to scale as 
\begin{equation}
\xi_L \sim (T - T_{\rm c})^{-\nu} ,
\label{eq:xiL-scale}
\end{equation}
where $\nu$ is the critical exponent for the correlation length. This scaling
behavior is expected to also be valid for zero-temperature transitions when
the ground state is not degenerate. Since in this work we want to study the
thermodynamic limit at finite but nonzero temperatures, we postulate that 
the scaling ansatz in Eq.~(\ref{eq:xiL-scale}) also holds for the model 
with a bimodal disorder distribution when $T > 0$.

The standard spin-glass susceptibility $\chi_{\rm SG} = \chi_{\rm SG}({\bf 
k} = 0)$ can also be defined via $\chi_{\rm SG} = N[\langle 
q^2\rangle]_{\rm av}$, where
\begin{equation}
q = \frac{1}{N}\sum_{i = 1}^N S_i^{a} S_i^{b} .
\label{eq:q}
\end{equation}
In Eq.~(\ref{eq:q}) $\{S_i^a\}$ and $\{S_i^b\}$ are two copies of 
the system with 
the same disorder. According to finite-size scaling we expect that
\begin{equation}
\chi_{\rm SG} \sim (T - T_{\rm c})^{-\gamma} ,
\label{eq:chi-scale}
\end{equation}
and at criticality
$\chi_{\rm SG}(T = T_{\rm c}) \sim L^{2 - \eta}$, where $\eta$
is the anomalous dimension exponent of the correlation function $G(r)$,
\begin{equation}
G(r,T) = [\langle S_i S_{i+r}\rangle^2]_{\rm av} = \frac{1}{r^{d-2+\eta}}e^{-r/\xi(T)} .
\label{eq:corr}
\end{equation}

In addition, we study the dimensionless Binder ratio\cite{binder:81}
defined via
\begin{equation}
g = \frac{1}{2}\left[3 - 
\frac{[\langle q^4 \rangle]_{\rm av}}{[\langle q^2 \rangle]_{\rm av}^2}
\right] .
\label{eq:g}
\end{equation}
In Sec.~\ref{sec:fss} we plot the Binder ratio as a function of the 
correlation length divided by the system size. The method has the advantage 
that if data for different disorder distributions lie on the same universal 
curve, the systems are in the same universality class.\cite{katzgraber:06} 
Finally, we also compute the specific heat of the 
system,\cite{comment:specific}
\begin{equation}
C_{\rm V} = \frac{1}{T^2} [\langle {\mathcal H}^2 \rangle - 
\langle {\mathcal H}\rangle^2]_{\rm av} ,
\label{eq:cv}
\end{equation}
which is expected to scale as 
\begin{equation}
C_{\rm V} \sim (T - T_{\rm c})^{-\alpha} .
\label{eq:cv-scale}
\end{equation}
For zero-temperature transitions the critical contribution of
the specific heat can also be written as $C_{\rm V}(T) \sim
T^{d\nu}$ using zero-transition-temperature scaling
relations.\cite{baker:75,campbell:04}

If two systems are in the same universality class they share
identical values of the critical exponents, as well as the
values of different observables at criticality [e.g., $g(T_{\rm
c})$].\cite{comment:bc} Assuming the power-law behaviors for the
different observables [Eqs.~(\ref{eq:xiL-scale}), (\ref{eq:chi-scale}),
and (\ref{eq:cv-scale})] at low but nonzero temperature, we study
the values of the effective critical exponents.

\section{Summary of previous results}
\label{sec:previous}

The critical properties of the two-dimensional Ising spin glass
with Gaussian-distributed interactions are firmly established
from zero-temperature simulations. Because the ground state is not
degenerate, the correlation function $G(r,T) = 1$ for all $r$ at $T =
T_{\rm c} = 0$.  Therefore, by definition, $\eta = 0$, $\xi(T = 0)
= \infty$, and for the Binder ratio at zero temperature $g(T = 0) =
1$. It is now well established from domain-wall measurements at $T =
0$, confirmed by size-dependent ground-state energy measurements
\cite{campbell:04} that $\theta = -0.282(2)$, hence the thermal
exponent $\nu$ (whose value is not fixed by the unique ground-state
condition) is $\nu \equiv -1/\theta = 3.54(2)$. Measurements of $\nu$
at finite temperatures via Monte Carlo simulations give consistent
estimates,\cite{katzgraber:04} yet only if large enough system sizes
are simulated. The different expected critical exponents at zero temperature
are summarized in Table \ref{tab:exp}.

For the two-dimensional Ising spin glass with bimodal-distributed 
interactions the situation is, however, much more complicated because of the 
highly degenerate ground state, as well as the quantized energy spectrum. 
As noted for instance in Ref.~\onlinecite{lukic:04}, the na\"{i}ve 
prediction for the low-temperature limit specific heat for the 
two-dimensional system of size $L$ and gap $A = 4$ is
\begin {equation}
C_{\rm V}(T)= \frac{16}{(TL)^2}\frac{N_1(L)}{N_0(L)}e^{-4J/T} .
\label{eq:naive}
\end{equation}
{\it A priori} this low-temperature exponential finite-size behavior 
should always hold for $T \ll T^{*}(L)$. Surprisingly, there has been a 
longstanding controversy concerning the low-temperature behavior of the 
specific heat in the thermodynamic limit. An exponential scaling of the 
free energy, and thus correspondingly of all thermodynamic quantities, has 
been first proposed by Wang and Swendsen.\cite{wang:88} They surmised that
\begin{equation}
C_{\rm V} \sim \frac{1}{T^P} e^{-AJ/T } \; .
\label{wangCv}
\end{equation}
The numerical parameters $A = 4$ and $P = 2$ can be expected from the 
aforementioned arguments regarding the gap in the excitation spectrum.
In addition, according to hyperscaling, the singular part of the 
free energy scales as $\xi^{-d}$ with $d = 2$ and so, if $C_{\rm V}$ 
scales exponentially, we expect that the correlation length $\xi$ scales 
as
\begin{equation}
\xi \sim e^{nJ/T} ,
\label{eq:expscale}
\end{equation}
with $n = A/2$, as predicted first by Saul and Kardar.\cite{saul:93} 

\begin{table}
\caption{
Critical exponents for both disorder distributions at zero
temperature. For the bimodal disorder distribution, exponential
scaling is expected, i.e., $\nu = \infty$ (see the main text for
details). The critical exponent $\eta$ in the bimodal case follows
from Refs.~\onlinecite{mcmillan:83}, \onlinecite{bhatt:88},
\onlinecite{wang:88}, \onlinecite{saul:93}, \onlinecite{blackman:98},
and \onlinecite{amoruso:06a} and are estimated by extrapolating
finite-temperature data to $T = 0$. In the Gaussian case the estimate
for $\nu$ is from Ref.~\onlinecite{hartmann:01a}. The remaining
critical exponents can be computed from the zero-temperature scaling
relations $\alpha = -d \nu$ and $\gamma = \nu (d - \eta)$, where 
$d = 2$ is the space dimension.
\label{tab:exp}
}
\begin{tabular*}{\columnwidth}{@{\extracolsep{\fill}} l l l }
\hline
\hline
Disorder &  $\nu$  & $\eta$ \\
\hline
Gaussian & $3.54(2)$ & $0$               \\
Bimodal   & $\infty$  & $0.14$ -- $0.40$  \\
\hline
\hline
\end{tabular*}
\end{table}

Wang and Swendsen\cite{wang:88} calculated numerically the
specific heat of the model. Surprisingly, they found $A = 2$,
thus suggesting a nontrivial scaling of the free energy. However,
their measurements were restricted to small system sizes and to few
disorder realizations, and their results indicated strong corrections
to scaling. These conclusions stand in contrast to those from work by
Saul and Kardar,\cite{saul:93} who argue that $A = 4$. In addition,
these authors also estimated $n = 2$, a behavior which appeared to
be confirmed independently in work by Houdayer\cite{houdayer:01} who
studied the finite-size scaling of the Binder ratio\cite{binder:81}
as well as by Katzgraber {\em et al.}\cite{katzgraber:05b} who
were the first to study the finite-size scaling of the finite-size
correlation length directly via Monte Carlo simulations. Note that an
exponential scaling of the correlation length (for all $A$) implies
that the effective critical exponent $\nu$ is infinite. The Wang and
Swendsen\cite{wang:88} value for $A$ ($A=2$) was strongly supported by
the numerical work of Lukic {\em et al.}\cite{lukic:04} who computed
the specific heat of the model for intermediate system sizes ($L \le
50$), using Pfaffian matrix algebra techniques. From their analysis
they concluded that $A$ could be estimated very accurately and that
$C_{\rm V}(T)$ tends to the functional form in Eq.~(\ref{wangCv}) with
$P=2$ and $A = 2.02(3)$,\cite{comment:katzgraber} in agreement with the
results of Ref.~\onlinecite{wang:88}. Recently, a new scenario has been
proposed in Ref.~\onlinecite{joerg:06a}: While at zero temperature the
model with bimodally distributed couplings still exhibits exponential
scaling with $A=4$, at finite but low temperatures in the thermodynamic
limit the two-dimensional Ising spin glass with bimodal couplings falls
into the {\em same} universality class as the system with Gaussian-distributed 
disorder. For nonzero temperatures and in the thermodynamic
limit the observables are claimed to display power-law singularities
with the {\em same} critical exponents as the system with Gaussian
disorder. In this work we compute effective critical exponents with
systems larger than in Ref.~\onlinecite{joerg:06a} in the temperature
range where power-law scaling is expected to occur\cite{joerg:06a}
($0.2 \lesssim T \lesssim 0.5$ for $L \gtrsim 50$) and show that
if the observables can be interpreted to exhibit power-law scaling,
the critical exponents for the system with bimodally distributed
couplings cannot be estimated reliably\cite{comment:othermodels}
from the system sizes studied in Ref.~\onlinecite{joerg:06a}.

\section{Results}
\label{sec:crit}

We present numerical data on various observables, comparing where
appropriate the systems with bimodal and Gaussian disorder. We
note that for the model with bimodal disorder in addition to the
crossover temperature $T^{*}(L)$ due to the energy gap, there is a
finite-size crossover temperature $T_{\xi}(L)$ fixed by the condition
$L \sim \xi(T)$.  Above $T_{\xi}(L)$, the observables are close to the
thermodynamic limit values while below $T_{\xi}(L)$, $\chi_{\rm SG}$
and $\xi_L$ tend to be size-limited and thus temperature independent.
For all system sizes studied we find $T_{\xi}(L) > T^{*}(L)$,
meaning that as $T$ is lowered the size limited condition on $\xi_L$
and $\chi_{\rm SG}$ sets in well before the effect of the gap. The
specific heat is approximately size independent down to $T^{*}$,
but is strongly affected by the gap: below $T^{*}(L)$ the specific
heat drops exponentially with decreasing $T$.

We now test the hypothesis of Ref.~\onlinecite{joerg:06a} that the
effective finite-temperature exponents $\eta_{\rm eff}$ and $\nu_{\rm
eff}$ exist and are identical to the values with Gaussian disorder in
$d = 2$, which are $\eta = 0$ and $\nu = 3.54(2)$, respectively. These
values imply that the exponent of the specific heat is $\alpha \approx
-7.1$ ($\alpha =  -d\nu$) and for the susceptibility exponent $\gamma
\approx 7.1$ [$\gamma = \nu(d - \eta)$].

\subsection{Specific heat}

\begin{figure}
\includegraphics[width=9.0cm]{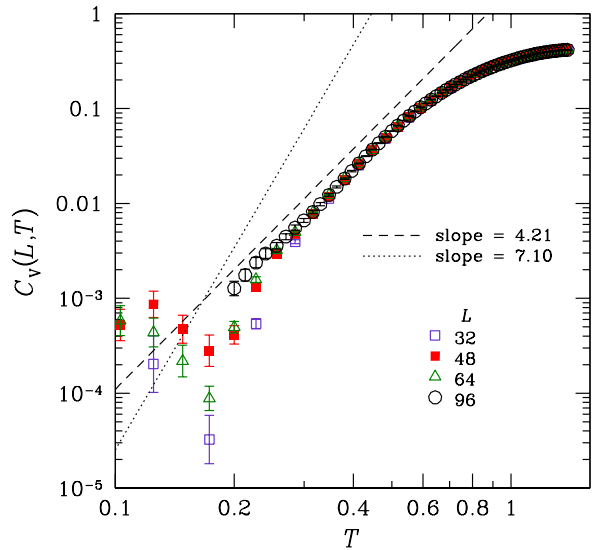}
\vspace*{-1.0cm}
\caption{
(Color online)
Log-log plot of the specific heat $C_{\rm V}$ vs temperature $T$ for 
several system sizes up to $L = 96$. For $0.2 \lesssim T \lesssim 0.5$ 
the data seem to be approximately independent of system size and follow 
a power-law behavior, i.e., $C_{\rm V} \sim T^{-\alpha}$ with 
$\alpha \approx -4.21(2)$ (the dashed line with slope $4.21$
is a guide to the eye). 
For the system sizes studied, the data seem incompatible with a 
low-temperature, large-size-limiting 
effective exponent $\alpha$ equal to the Gaussian estimate for the critical 
exponent $\alpha \approx  -7.1$ (dotted line with slope $7.1$).
The fluctuations at $T \lesssim 0.2$ can be ascribed to the exponential
behavior at low enough temperatures.
}
\label{fig:cv_vs_T-loglog}
\end{figure}

\begin{figure}
\includegraphics[width=9.0cm]{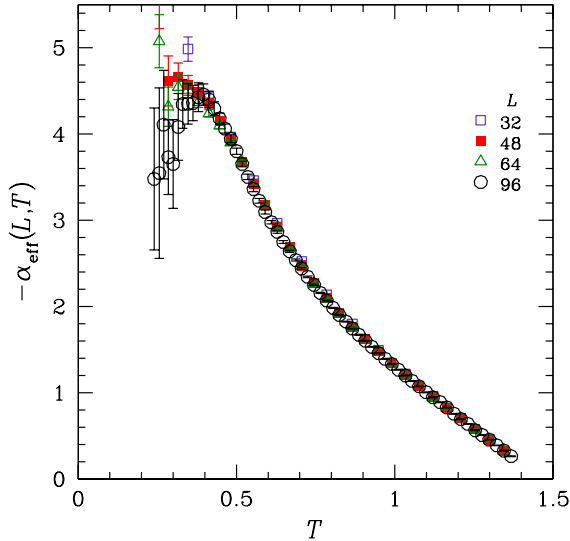}
\vspace*{-1.0cm}
\caption{
(Color online)
Effective exponent $\alpha_{\rm eff}(L,T) = -d\ln[C_{\rm V}(L,T)]/d\ln[T]$
as a function of temperature for different system sizes $L$.
For the system sizes shown, the data cannot be extrapolated in a reliable way
to $\alpha_{\rm eff} = -7.1$, the value obtained for the model with 
Gaussian-disordered bonds. Therefore it is rather difficult to test if
the effective critical exponents agree with the known Gaussian values.
}
\label{fig:dlog_cv_dlog_T}
\end{figure}

Figure \ref{fig:cv_vs_T-loglog} shows a log-log plot of the specific
heat $C_{\rm V}(L,T)$ as a function of $T$.  The numerical results
are consistent with those of the analogous plot shown in Fig.~4 
of Ref.~\onlinecite{joerg:06a}, but the present data extend to
$L = 96$. The slope in the range $0.2 \lesssim T \lesssim 0.5$
is $-d\ln[C_{\rm V}(L,T)]/d\ln[T] \sim -4.21(2)$, i.e., $\alpha
\approx -4.21$.  We also perform a point-by-point differential of
the data for all $T$ (using a second-order midpoint differentiation
combined with a bootstrap analysis to estimate the error bars)
and thus estimate the effective exponent $\alpha_{\rm eff}(L,T)$
as a function of temperature.  Figure \ref{fig:dlog_cv_dlog_T}
displays $\alpha_{\rm eff}(L,T) = -d\ln[C_{\rm V}(L,T)]/d\ln[T]$
as a function of $T$. $\alpha_{\rm eff}(L,T)$ should tend to the
thermodynamic critical exponent $\alpha_{\rm eff}$ in the limit $L
\rightarrow \infty$ followed by $T \rightarrow 0$.  An extrapolation
of the data for $T > T^{*}(L)$, where $-d\ln[C_{\rm V}(L,T)]/d\ln[T]$
is independent of system size $L$, cannot be performed in a reliable
way to test if the effective exponent agrees with the expected Gaussian
value of $\alpha_{\rm eff} \approx -7.1$.

For $0.2 \lesssim T \lesssim 0.5$ the data seem to saturate (see
Fig.~\ref{fig:dlog_cv_dlog_T}) although strong fluctuations are
present.  This ``plateau region'' resembles the behavior predicted
by Fisch\cite{fisch:06} who argues that there should be a rather
broad region in temperature just above $T^{*}(L)$ for large (but not
infinite) $L$, where $C_{\rm V} \sim T^x$ with $x = 5.25(20)$. This
is equivalent to a temperature-independent $\alpha_{\rm eff} =
-5.25$. Given the uncertainties in the estimate of the exponent,
the agreement between the prediction by Fisch and the Monte Carlo
data presented here is reasonably good.  For the system sizes studied,
which are larger than the ones studied in Ref.~\onlinecite{joerg:06a},
the exponent $\alpha_{\rm eff}$ in this large-$L$ low-$T$ region seems
to be different from the Gaussian critical exponent $\alpha = -2\nu
\approx  -7.1$. (In fact, in the Gaussian model the low-temperature
specific heat is dominated by noncritical contributions\cite{lukic:04}
and the true critical behavior is not directly visible). Therefore, if
both models share the same universality class, the system with bimodal
disorder displays huge corrections to scaling and thus simulations
at {\em considerably} larger system sizes would be required to prove
this beyond any reasonable doubt.

\subsection{Correlation length and susceptibility}

\begin{figure}
\includegraphics[width=9.0cm]{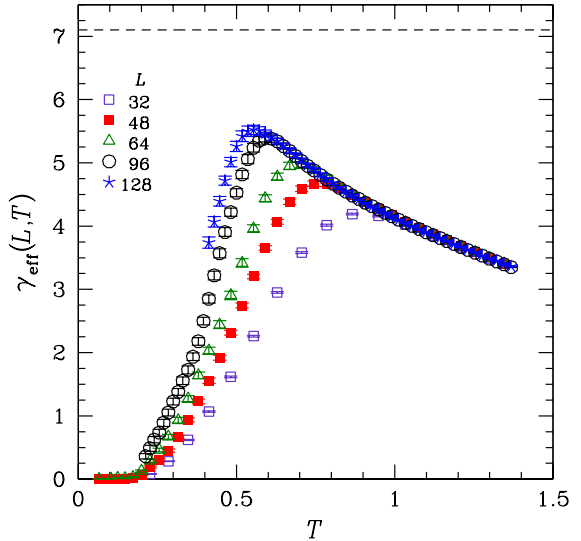}
\vspace*{-1.0cm}
\caption{
(Color online)
Effective exponent $\gamma_{\rm eff}(L,T) = -d\ln[\chi_{\rm 
SG}(L,T)]/d\ln[T]$ as a function of temperature for different system 
sizes $L$. An extrapolation to the low-temperature regime is difficult 
with system sizes limited to $L \le 128$. The horizontal dashed line
corresponds to the expected Gaussian value $\gamma \approx 7.1$. An 
agreement or disagreement with $\gamma \approx 7.1$ cannot be ruled out. 
}
\label{fig:dlog_chisg_dlog_T}
\end{figure}

\begin{figure}
\includegraphics[width=9.0cm]{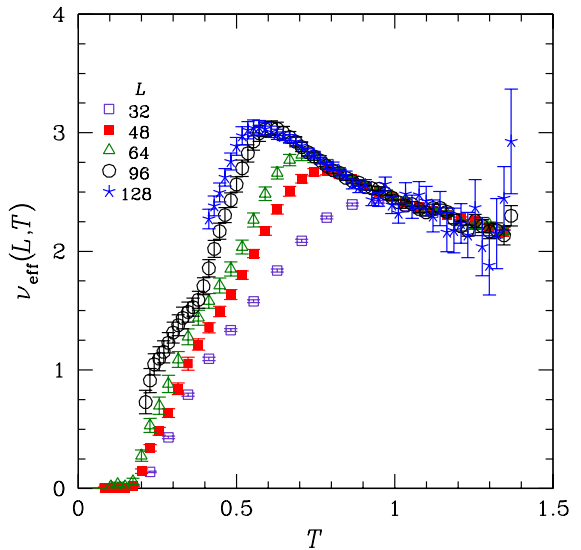}
\vspace*{-1.0cm}
\caption{
(Color online)
Effective exponent $\nu_{\rm eff}(L,T) = -d\ln[\xi(L,T)]/d\ln[T]$ as a 
function of temperature for different system sizes $L$. An extrapolation 
to the low-temperature regime is difficult with system sizes limited to 
$L \le 128$. Any extrapolation to the low-$T$ behavior where $\nu \approx
3.54$ would be difficult to perform with the current data.
}
\label{fig:dlog_xi_dlog_T}
\end{figure}

From the assumed power-law critical behavior of the spin-glass
susceptibility one can define an effective exponent $\gamma_{\rm
eff}(L,T) = -d\ln[\chi_{\rm SG}(L,T)]/d\ln[T]$ using point-by-point
differentiation. Figure \ref{fig:dlog_chisg_dlog_T} shows
$\gamma_{\rm eff}(L,T)$ against $T$. The thermodynamic limit
domain where the effective exponent is size independent can be
seen clearly; for each $L$ the data peel off the thermodynamic
limit line at the finite-size limited $T_{\xi}(L)$. However, even
with data up to $L = 128$ there is no reliable way to extrapolate
to the critical value at infinite $L$ and $T$ tending to zero. A
similar conclusion can be reached for $\nu_{\rm eff}(L,T) =
-d\ln[\xi(L,T)]/d\ln[T]$, Fig.~\ref{fig:dlog_xi_dlog_T}.  The error
bars in Figs.~\ref{fig:dlog_cv_dlog_T}, \ref{fig:dlog_chisg_dlog_T},
and \ref{fig:dlog_xi_dlog_T} have been calculated via a bootstrap
estimate.\cite{efron:94}

In Ref.~\onlinecite{joerg:06a} the Caracciolo finite-size scaling
technique\cite{caracciolo:95,palassini:99b} was used to extrapolate
$\xi(L,T)$ and $\chi_{\rm SG}(L,T)$ data for the $\pm J$ model
towards infinite size, in order to extract the exponent $\eta$
from the critical scaling relation $\chi_{\rm SG}(L,T) \sim
\xi(L,T)^{2-\eta}$. The data are interpreted as showing that $\eta
\sim 0$.\cite{joerg:06a}

\begin{figure}
\includegraphics[width=9.0cm]{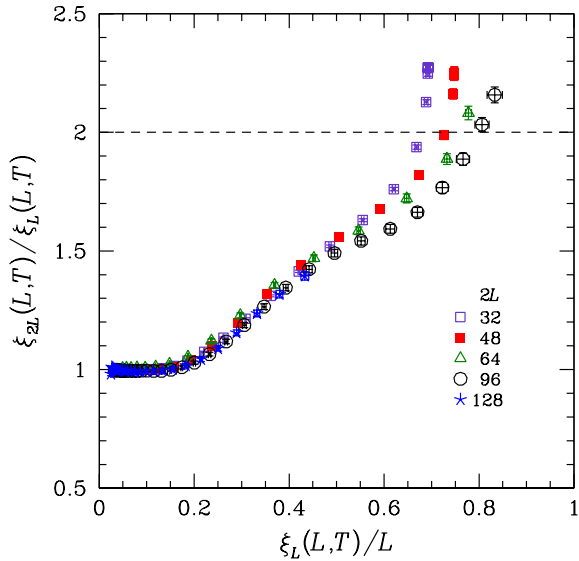}
\vspace*{-1.0cm}
\caption{
(Color online)
$\xi(2L,T)/\xi(L,T)$ against $\xi(L,T)/L$. If corrections to scaling were
small, all data should lie on an universal curve. This is a 
prerequisite to perform an extrapolation to the bulk regime. Clearly, the 
data do not scale well.
}
\label{fig:xi2L_ov_xiL_vs_xiLovL}
\end{figure}

\begin{figure}
\includegraphics[width=9.0cm]{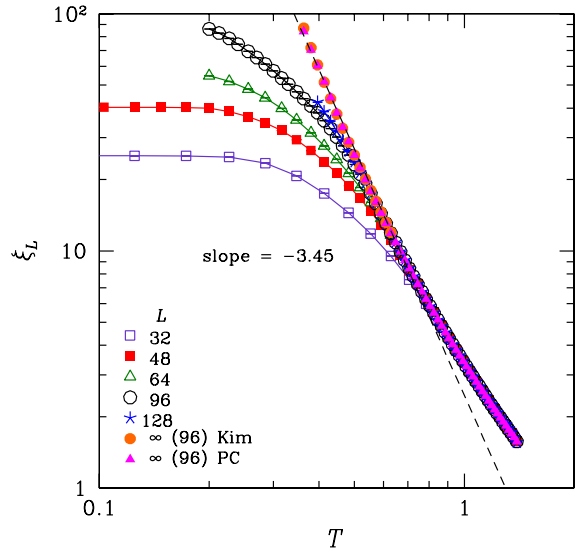}
\vspace*{-1.0cm}
\caption{
(Color online) Log-log plot of the finite-size correlation length
$\xi(L,T)$ as a function of temperature for several system sizes. The
solid orange circles represent extrapolated data to the thermodynamic
limit from $L = 96$ using the method of Kim\cite{kim:94} and the solid
pink triangles represent the data extrapolated to $L = \infty$ using
the method of Palassini and Caracciolo (PC).\cite{palassini:99b} 
The (extrapolated) data
seem to follow a power-law behavior with $\nu \approx 3.45$, which is
close to the value of the critical exponent for Gaussian-distributed
disorder, $\nu = 3.54(2)$.  The dashed line is a guide to the eye.
}
\label{fig:xi_vs_T-loglog}
\end{figure}

This technique assumes that there are well-behaved scaling functions
for $\xi(L,T)$ and $\chi_{\rm SG}(L,T)$, such that, in particular,
$\xi(2L,T)/\xi(L,T)$ is a unique function of $\xi(L,T)/L$. Otherwise
the Caracciolo scaling procedure cannot be applied as systematic errors
would be introduced. Figure \ref{fig:xi2L_ov_xiL_vs_xiLovL} shows
data for $\xi(2L,T)/\xi(L,T)$ vs $\xi(L,T)/L$ with $L$ between $16$
and $48$. It can be seen that the finite-size corrections to scaling
are strong because the curves for different $L$ do not superimpose,
and also because the ratios $\xi(2L,T)/\xi(L,T)$ increase beyond
a value of 2 at low temperatures (in the absence of corrections 2
is the strict $T_{\rm c}$ limit of the ratio). This means that the
Caracciolo procedure (or any similar protocol such as the one by
Kim\cite{kim:94} used by Katzgraber {\em et al.}\cite{katzgraber:04}
for the model with Gaussian disorder) has to be performed with
considerable care for bimodally distributed disorder in two space
dimensions.  In Fig.~\ref{fig:xi_vs_T-loglog} we show a log-log
plot of the correlation length as a function of temperature together
with an extrapolation using the methods of Kim\cite{kim:94} as well
as Palassini and Caraciolo.\cite{caracciolo:95,palassini:99b} Both
extrapolations agree very well.  While the extrapolated data seem to
follow a power-law behavior with $\nu \approx 3.45$, the extrapolation
method is not reliable due to strong corrections to scaling (see 
Fig.~\ref{fig:xi2L_ov_xiL_vs_xiLovL}).  Until data on much larger
system sizes become available it does not seem plausible to give
a reliable account of the large-$L$, low-$T$ limiting functional
behavior of $\chi_{\rm SG}(L,T)$ or $\xi(L,T)$.

\begin{figure}
\includegraphics[width=9.0cm]{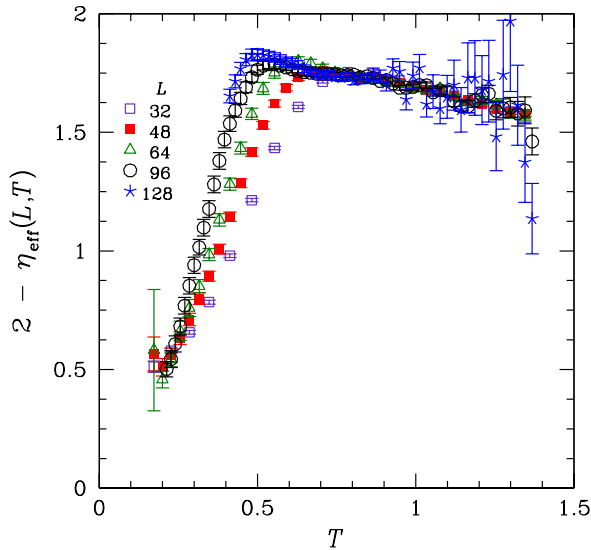}
\vspace*{-1.0cm}
\caption{
(Color online)
Effective exponent $2 - \eta_{\rm eff}(L,T) = d\ln[\chi_{\rm 
SG}(L,T)]/d\ln[\xi(L,T)]$ as a function of temperature for different 
system sizes $L$. For all system sizes and temperatures studied 
$\eta_{\rm eff}$ is always greater than $0.2$, although an extrapolation to
$\eta_{\rm eff} = 0$ cannot be ruled out.
}
\label{fig:dlog_chisg_dlog_xi}
\end{figure}

Alternatively, with no extrapolation, one can define an effective exponent
\begin{equation}
2 - \eta_{\rm eff}(L,T) = -\frac{d\ln[\chi_{\rm SG}(L,T)]}{d\ln[\xi(L,T)]}
\end{equation}
via differentiation of the data. For the available system sizes $\eta_{\rm 
eff}(L,T)$ is always greater than $\sim 0.2$ (see
Fig.~\ref{fig:dlog_chisg_dlog_xi}). A reliable extrapolation to infinite 
$L$ and $T$ tending to zero would again require data of much larger system 
sizes.

There have been numerous estimates of $\eta$ at zero temperature
for the two-dimensional Ising spin glass with bimodally
distributed disorder, including direct measurements of the
correlation function $G(r)$ by combinatorial or Monte Carlo
methods,\cite{mcmillan:83,blackman:98} and there is a general consensus
that $\eta \gtrsim 0.15$. Indeed it can be noted that the finite-$T$
raw $G(r)$ data in McMillan's\cite{mcmillan:83} Fig.~1 are by
inspection incompatible with $\eta = 0$.

From the definition of $\eta$ through $G(r)$, when there is a
degenerate ground state at $T = 0$, the time average spin-spin
correlation function must decay with increasing distance $r$ implying
a positive-definite value for $\eta$.  {\it A fortiori} at any $T$
slightly above zero one would expect $G(r,T) \le G(r,0)$ for all $r$
except in quite exceptional cases.  Otherwise, a limiting $\eta_{\rm
eff} = 0$ [meaning $G(r,T)=1$ for all $r$ at $T$ close to zero]
appears to be ruled out from basic physical principles.

\subsection{Gaussian disorder}

\begin{figure}
\includegraphics[width=9.0cm]{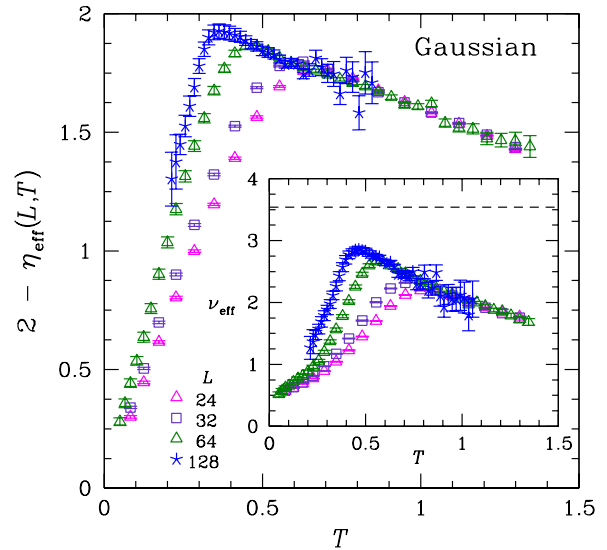}
\vspace*{-1.0cm}
\caption{
(Color online)
Effective exponent $2 - \eta_{\rm eff}(L,T) = d\ln[\chi_{\rm
SG}(L,T)]/d\ln[\xi(L,T)]$ as a function of temperature for different
system sizes $L$ for Gaussian disorder. The data extrapolate well to
$\eta \approx 0$. Note that for $L = 128$ the data for $T \gtrsim 0.9$
have been dropped due to strong fluctuations. Inset: Effective exponent
$\nu_{\rm eff}$ as a function of temperature for different system sizes
$L$. The dashed line corresponds to the zero-temperature estimate
from the stiffness exponent, $\nu = 3.54(2)$. 
}
\label{fig:eta_nu--gauss}
\end{figure}

\begin{figure}
\includegraphics[width=9.0cm]{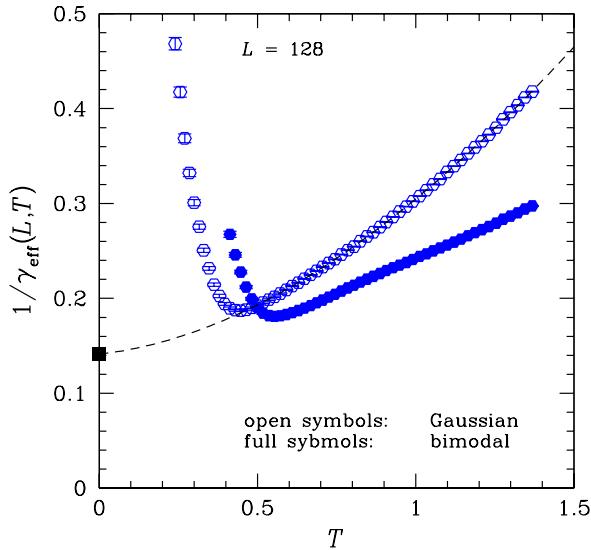}
\vspace*{-1.0cm}
\caption{
(Color online)
Effective critical exponent $1/\gamma_{\rm eff}$ for both Gaussian (open
symbols) and bimodal disorder (full symbols) as a function of temperature $T$.
While in the Gaussian case the $T = 0$ limit is known ($\gamma^{-1} \approx
0.14$, black square) and thus the extrapolation from the finite-temperature 
data can be performed via a simple third-order polynomial (dashed line), 
the data for the bimodal case can be extrapolated to any arbitrary value 
including $1/\gamma_{\rm eff} = 0$, which corresponds to exponential scaling.
}
\label{fig:inv_gamma}
\end{figure}

We have also computed the effective critical exponents for the
correlation length [$\nu_{\rm eff}(L,T)$], the specific heat
[$\alpha_{\rm eff}(L,T)$], the susceptibility [$\gamma_{\rm
eff}(L,T)$], and correlation function [$\eta_{\rm eff}(L,T)$] for the
two-dimensional Ising spin glass with Gaussian-distributed disorder
in order to test corrections to scaling in that model. In this case
\begin{equation}
P(J_{ij}) \sim e^{-J_{ij}^2/2J}
\end{equation}
in Eq.~(\ref{eq:ham}).  The results are qualitatively similar to the
results found for the model with bimodally distributed disorder, but
the data extend to lower temperatures thus making an extrapolation to
zero temperature slightly more reliable. Still, without the knowledge
of the zero-temperature estimate of the stiffness exponent $\theta$,
$\nu = -1/\theta$ could not be determined to such high precision.
In Fig.~\ref{fig:eta_nu--gauss} we illustrate this behavior with data
for $\eta_{\rm eff}$ and $\nu_{\rm eff}$ as a function of temperature.
In Fig.~\ref{fig:inv_gamma} we compare the effective critical exponent
$\gamma_{\rm eff}^{-1}$ for Gaussian and bimodal disorder for $L =
128$. While the extrapolation in the Gaussian case can be done easily
to $T = 0$, since the zero-temperature limit is well known, this is
difficult for the bimodal case.

\section{Universality and finite-size scaling}
\label{sec:fss}

The critical values of the correlation length divided by the system
size $\xi_L(T_{\rm c})/L$ [Eq.~(\ref{eq:xiL})] and the Binder
ratio $g(T_{\rm c})$ [Eq.~(\ref{eq:g})] are characteristic of the
universality class of a continuous transition. These are linked
to $\eta$ at criticality because they represent various ratios
of integrals with $G(r) \sim r^{d-2+\eta}$. For example, in a
strip geometry for two space dimensions, $\xi_L/L = 1/(\pi\eta)$
at criticality.\cite{cardy:84} The correlation length divided
by system size of the two-dimensional spin glass with Gaussian-distributed 
disorder diverges for $T \rightarrow 0$ and the Binder
ratio tends to $1$. If the two-dimensional Ising spin glass with
bimodal interactions lies in the same universality class as the system
with Gaussian disorder, then identical values for these parameters
at criticality should be observed.  The Binder ratio values $g(L,T)$
become temperature independent within the error bars for $T < T^{*}(L)$
providing estimates of the zero-temperature values $g(L,0)$. There
are corrections to scaling but the series of points appear to tend
to a large-$L$ limit which is significantly less than unity.

\begin{figure}
\includegraphics[width=9.0cm]{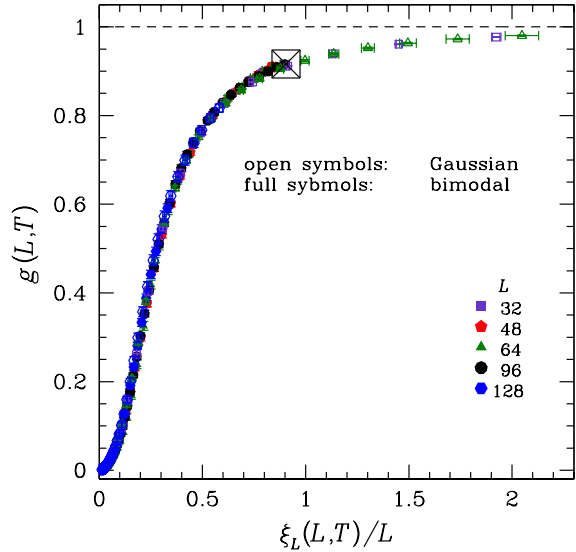}
\vspace*{-1.0cm}
\caption{
(Color online)
Binder ratio $g(L,T)$ as a function of the finite-size correlation length
divided by the system size, $\xi_L(L,T)/L$. While the data for Gaussian
disorder seem to extrapolate to $g(L,0) = 1$ and $\xi_L(L,T)/L = \infty$, the
data for bimodal disorder seem to stop at $g(L,0) \approx 0.92$ and 
$\xi_L(L,T)/L \approx 0.91$ (marked by a boxed cross).
}
\label{fig:g_vs_xi}
\end{figure}

For any continuous transition, at large $L$ and $T$ approaching
$T_{\rm c}$ the Binder ratio is a nontrivial function of the variable
$\xi_L/L$.\cite{kim:96,katzgraber:06,joerg:06} A plot of $g(L,T)$ vs
$\xi_L(L,T)/L$ (Fig.~\ref{fig:g_vs_xi}) shows a remarkable behavior:
all points for both bimodal and Gaussian disorder are on a unique
curve. This is particularly striking as the data span both the
regions $T > T_{\xi}(L)$ and $T < T_{\xi}(L)$.  There is, however,
a qualitative difference between the Gaussian data and the bimodal
data. For the former at each $L$ the data points extend to the same
zero-temperature end point $\xi_L(L,T = 0)/L = \infty$ and $g(L,T =
0) = 1$, while for the system with bimodal disorder the end points
for different $L$ seem to cluster and not grow beyond $\xi_L(L,T =
0)/L \sim 0.91(2)$ and $g(L,T = 0) \sim 0.92(2)$. In fact, the data do
grow slightly, but the growth rate is within statistical error bars.
This point can be interpreted as the zero-temperature critical
parameters for this model. Thus, while the scaling functions agree,
at zero temperature both models seem not to be in the same universality
class.  An alternative explanation could be that the model with bimodal
disorder is ``marginal,'' i.e., the endpoint might approach $g(L,T =
0) = 1$ logarithmically slow. In the bulk regime, which corresponds
to the lower left corner of Fig.~\ref{fig:g_vs_xi}, data for $g(L,T)$
and $\xi_{L}(L,T)/L$ agree and thus suggest that {\em both models
might share a common finite-temperature universality class}. Note
that this cannot be inferred from studying the critical exponents
due to large corrections to scaling, as shown in Sec.~\ref{sec:crit}.

\section{Conclusion}
\label{sec:conclusions}

We have carried out a careful numerical study of the two-dimensional
Ising spin-glass model with bimodally-distributed disorder at finite
temperatures with system sizes up to $L=128$, measuring the specific
heat, finite-size correlation length, the spin-glass susceptibility,
and the Binder ratio. The data neither confirm nor disprove the
hypothesis\cite{joerg:06a} that in the large-$L$, low-but-nonzero-$T$
limit the effective critical exponents of the system are identical
to the known critical exponents of the two-dimensional Ising spin
glass with Gaussian-distributed disorder, although the data seem
to follow a power-law behavior at finite $T$ (see, for example,
Fig.~\ref{fig:cv_vs_T-loglog}).  Since the evidence for the critical
exponents of the two-dimensional Ising spin glass with bimodally
distributed disorder being identical to those of the pure Gaussian
case is weak, a further critical analysis of the estimates for the
other distributions studied in Ref.~\onlinecite{joerg:06} might
be opportune.  A plot of the Binder ratio against the finite-size
correlation length (Fig.~\ref{fig:g_vs_xi}), which shows very small
corrections to scaling, only suggests that in the bulk regime and
for finite temperatures both models might share a common universality
class. Corrections to scaling are extremely large in the bimodal case.
Therefore our main result is that simulations with larger system sizes
and lower temperatures are needed to conclusively compute the critical
exponents to prove the scenario proposed in Ref.~\onlinecite{joerg:06a}
in which both models with Gaussian and bimodally distributed disorder
are in different universality classes at $T = 0$, where the degeneracy
of the ground state plays a key role, yet the models share the same
universality class at finite nonzero temperatures and very large
system sizes.

\begin{acknowledgments} 

We would like to thank R.~Fisch, K.~Hukushima, T.~J\"org, J.~Poulter,
M.~Troyer, and A.~P.~Young for fruitful discussions, and T.~J\"org
for pointing out the usefulness of plotting the data in the way shown
in Fig.~\ref{fig:g_vs_xi}.  The simulations were performed on the
Hreidar and Gonzales clusters at ETH Z\"urich.

\end{acknowledgments}

\bibliography{refs,comments}

\begin{thebibliography}{39}
\bibitem[{com({\natexlab{a}})}]{comment:specific}
\bibinfo{note}{Note that the specific heat can also be computed as a derivative
  of the internal energy $E = - [\langle {\mathcal H} \rangle]_{\rm av}$, i.e.,
  $C_{V} = \partial E /\partial T$ (constant volume). We have also computed the
  specific heat using this method and the results agree perfectly with the
  results using Eq.~(\ref{eq:cv}).}
\bibitem[{com({\natexlab{b}})}]{comment:bc}
\bibinfo{note}{Numerically one can only access finite system sizes. In this
  case the boundary conditions (here periodic) as well as the shape anisotropy
  of the models compared have to be the same. Only then will the models be in
  the same universality class, provided that the critical exponents as well as
  the values of different observables at criticality are the same.}
\bibitem[{com({\natexlab{c}})}]{comment:katzgraber}
\bibinfo{note}{Similar results have been found via an alternate analysis in
  Ref.~\onlinecite{katzgraber:05e}, although it is unclear if an extrapolation
  from the high-temperature behavior to $T = 0$ is plausible.}
\end{thebibliography}

\end{document}